\documentclass[fleqn,twoside]{article}
\usepackage{espcrc2}

\usepackage{graphicx}

\newcommand{\PS}{{\rm PS}}
\newcommand{\fPS}{{\ensuremath f_{\PS}}}
\newcommand{\mPS}[1]{{\ensuremath m_{\PS}^{#1}}}
\newcommand{\Order}{{\ensuremath\cal O}}  
\newcommand{\be}{\begin{equation}}
\newcommand{\ee}{\end{equation}}
\newcommand{\bea}{\begin{eqnarray}}
\newcommand{\eea}{\end{eqnarray}}
\newcommand{\non}{\nonumber}
\newcommand{\Eq}[1]{Eq.~(\ref{#1})}
\newcommand{\Fig}[1]{Fig.~\ref{#1}}
\newcommand{\chpt}{$\chi$PT}

\newcommand{\X}{{\mathcal B}}

\title{The Pseudoscalar Decay Constant}

\author{S. V. Wright\address{Division of Theoretical Physics,
    Department of Mathematical Sciences, University of Liverpool,
    Liverpool L69 3BX, UK}\thanks{Current Address: Physics Division,
    Argonne National Laboratory, Argonne IL 60439, USA.}\\
  \vspace{3mm}
  {\em UKQCD Collaboration}\\
}
       
\begin{document}

\begin{abstract}
We discuss insights that may be drawn from our recent 2
flavour $\Order(a)$--improved Wilson quark simulations.  We
discuss the evidence of the onset of chiral logarithms in the pion
decay constant.  An overview is given of current extrapolation methods
and a modification of chiral perturbation theory is presented as an
approach for sensibly extrapolating to the physical quark masses.
\vspace{1pc}
\end{abstract}

\maketitle

\section{Introduction}
In a recent work~\cite{Allton:2004qq} we presented results from the
lightest UKQCD Wilson quark simulations for the mass of the singlet
pseudoscalar meson and the pseudoscalar (pion) decay constant. 
In this work we shall
expand upon the pion decay constant extrapolation details and suggest
a simplistic first attempt at an improved extrapolation method.

The motivation for this recent calculation of the pseudoscalar decay
constant, $\fPS$, is the expectation of observing behaviour predicted
by Chiral Perturbation Theory (\chpt) as the pseudoscalar mass is
lowered.  In particular evidence of the onset of chiral logs in
observables.
The lowest order chiral Lagrangian has non-analyticity resulting from
loop corrections introduced in the parameters.  This non-analytic
behaviour provides a good check that the lattice calculation is in the
regime where chiral perturbation theory is valid.

Computational constraints have forced us to work at finite lattice
spacing, however the choice was made to compare our results to the
continuum predictions of \chpt.
Whilst Chiral perturbation theory has been formulated for finite
lattice spacing~\cite{Aoki:2003yv,Bar:2003mh}, it is at the cost of
additional parameters.

\section{Calculating $\fPS$}
Our results for the pseudoscalar decay constant were extracted from UKQCD
simulations at three different quark masses: $\kappa = 0.1358$,
$0.1355$ and $0.1350$.
All lattices had $16^{3} \times 32$ volumes with $\beta = 5.2$ and
a non-perturbatively improved clover action.
Using quark propagators with sources on the time planes $t=0$, 7, 15
and 23 allowed an improvement in the statistics of the simulation.
As the value of $\mPS{} L \approx 4$ in each of the cases we would
expect some finite volume effects in our calculations.  These are
taken into consideration~\cite{Colangelo:2003hf} in the subsequent
analysis.

\begin{table}[htb]
  \caption{The raw lattice value of $a\fPS$ is  given by using the
    order $a$ improved expression $(1+b_A \mPS{})(af_A+c_A af_P)$ and
    we tabulate these two contributions.~\cite{Allton:2004qq}
  }
  \label{tab:fpiData}
  \begin{tabular}{lll} \hline
    $\kappa$ & $af_A$ &  $af_P$ \\  \hline
    0.1358 & 0.0829(26) &  0.1457(78) \\
    0.1355 &  0.1055(14) &  0.1835(44) \\
    0.1350 & 0.1336(11) &  0.2468(33) \\
    \hline
  \end{tabular}
\end{table}

\begin{figure}[htb]
  \includegraphics[width=50mm,angle=90]{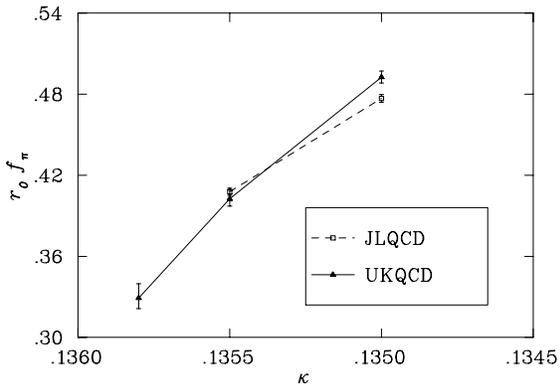}
  \caption[]{
    The pseudoscalar decay constant in units of $r_0$ from  UKQCD and
    JLQCD versus $\kappa$.\cite{Allton:2004qq}
  }
  \label{fig:ukvsjl}
\end{figure}

Our results for $\fPS$ were presented in~\cite{Allton:2004qq} and
are reproduced here in Table \ref{tab:fpiData} and \Fig{fig:ukvsjl}.  
The lattice value of $a\fPS$ used in \Fig{fig:ukvsjl} is found by
using the order $a$ improved expression $(1+b_A \mPS{})(af_{A}+c_{A}
af_P)$, where $f_{A,P}$ are the axial-vector and pseudoscalar decay
constants and their contributions are listed in Table \ref{tab:fpiData}.

To aid comparison with the only available comparable data, that of 
JLQCD~\cite{Aoki:2002uc}, we have used the same perturbative formulation of the
corrections and also the same prescription for $Z_{A}$, the
renormalisation factor.  Finally as there
is a discrepancy in the prescription for evaluating $r_{0}$ we apply
our determination of $r_{0}$ to their data.  The comparison of results
is shown in \Fig{fig:ukvsjl}.  The agreement at the lightest common $\kappa$
values is pleasing.
The exciting feature of our new results is the indication of curvature
versus $m_{q}$.
It is this result that has motivated the current investigation.

\section{Extrapolation Comparisons}
\label{sec:comparison}
The limitation of all current lattice calculation of the pseudoscalar
decay constant to large quark masses necessitates some form of
extrapolation to the physical quark (or equivalently pion) mass.  Over
the last few years it has become accepted within lattice groups that
some form of {\em chirally motivated} extrapolation is appropriate
\cite{Leinweber:1999ig,Leinweber:2001ac,Young:2002cj,Bernard:2002yk},
by using insights from chiral perturbation theory, when extrapolating
physical observables.  The situation is no different in the case of
$\fPS$, however opinions of what is {\em appropriate} differ between
groups.

The chiral perturbation prediction for the pion mass dependence of
$\fPS$ is~\cite{Colangelo:2003hf,Bijnens:1998fm}:
\bea
\frac{\fPS}{f_{\pi}(0)} & = &1 - 2 \left( 
      \frac{\mPS{}}{4\pi f_{\pi}(0)} \right)^{2}
      \log\left(\frac{\mPS{2}}{\Lambda_{4}^{2}}\right)\non\\
      && + \cdots
\label{eq:ChPT}
\eea
where $f_{\pi}(0)$ is the pseudoscalar (pion) decay constant in the
chiral limit.

\begin{figure}[htb]
  \includegraphics[width=55mm,angle=90]{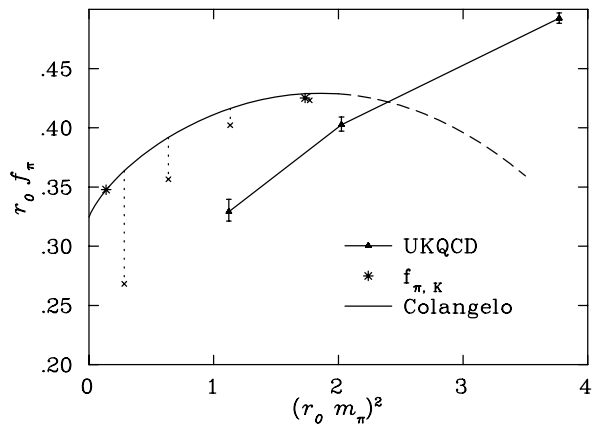}
  \caption[]{
   The pseudoscalar decay constant in units of $r_0$ from  UKQCD  
  versus the squared pseudoscalar meson mass.
   Also shown is an expression including chiral perturbation theory terms 
  to order $\mPS{4}$ which has been fitted (see ref.~\cite{Colangelo:2003hf}
  where we  use $\mu=0.75$ GeV and $\tilde{r}_F(\mu)=-2$) to agree with  the
  experimental values of $f_{\pi}$ and $f_K$ which are shown (*). An
  estimate~\cite{Colangelo:2003hf}  of the finite size effect expected from
  chiral perturbation theory (to order $\mPS{2}$) is shown by the vertical
  lines.
  }
  \label{fig:chlog}
\end{figure}%

Recent work~\cite{Colangelo:2003hf} has provided a way of estimating
the expected finite size effects.
The quark loops generate logarithmic corrections, but moreover they
are the source of the finite size effects.  
Using $L=1.5$ fm and $m_{\pi}=400$ MeV, which are close to our values
a suggestion of the size of the corrections may be determined.
The solid curve shown is an expression including chiral perturbation
theory terms to order $\mPS{4}$ which has been fitted (see
Ref.~\cite{Colangelo:2003hf} where we  use $\mu=0.75$ GeV and
$\tilde{r}_F(\mu)=-2$) to agree with the experimental values of
$f_{\pi}$ and $f_K$ which are shown (*). An
estimate~\cite{Colangelo:2003hf}  of the finite size effect expected
from chiral perturbation theory (to order $\mPS{2}$) is shown by the
vertical lines. 
As an experiment, we also continued the \chpt{} prediction for
$\fPS$ to almost twice the kaon mass (dashed curve).
It is clear the lattice simulation behaviour in this regime is
fundamentally different from the na\"{\i}ve predictions of \chpt.
This discrepancy indicates \chpt{} (as formulated) has a limited range
of overlap with current simulations.

\subsubsection*{JLQCD}
Recent progress in the extension of \chpt{} has been attempted by the
JLQCD collaboration~\cite{Aoki:2002uc}.  Their approach used \chpt{}
as motivation for their functional form which, in the dynamical case
($\kappa_{\rm sea}~=~\kappa_{\rm val}$) may be written as:
\be
\fPS = A + B \mPS{2} + C \mPS{4}
\label{eq:JLQCD}
\ee

\begin{figure}[htb]%
  \includegraphics[width=70mm]{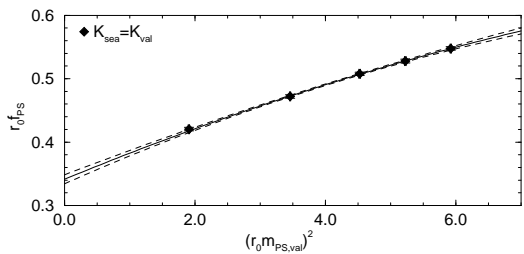}%
  \caption{Chiral extrapolation of the pseudoscalar meson decay
  constant. The fit is made using \Eq{eq:JLQCD}.~\cite{Aoki:2002uc}}
  \label{fig:JLQCD}
\end{figure}

The analytic (even powers of $\mPS{}$) terms of \Eq{eq:JLQCD} are
consistent with both \chpt{} and the JLQCD data set, as seen in
\Fig{fig:JLQCD}.  The success of such a fit may be seen to vindicate
``leaving the problem of the chiral
logarithm~[\ldots]~for future publication.''~\cite{Aoki:2002uc}

The problem with this approach is the selective nature of what parts
of \chpt{} are used.  The chiral logarithm term that occurs with a power of
the pseudoscalar mass less than $\mPS{4}$ is ignored.  Yet this is the
greatest contribution to the curvature in the small mass regime.  The
form is also inconsistent with the large mass behaviour of $\fPS$
which experience has shown to be proportional to $\mPS{2}$ up to
pseudoscalar masses of the order of 1.4 GeV${}^{2}$.  Thus, any
perceived successes of this form of extrapolation must be tempered by
the clear inconsistencies with the behaviour of QCD as established by
\chpt.

\subsubsection*{Dyson--Schwinger Equation}
An interesting alternative insight is that promoted by the
Dyson--Schwinger Equation (DSE) community.
The Dyson--Schwinger equations, whilst maintaining a Poincar\'{e}
covariant framework, provide a setting for a
non-perturbative chiral symmetry preserving truncation that
leads to an efficient one--parameter rainbow--ladder
model~\cite{Maris:1999nt}.  In this DSE approach, the pion is
well--understood and is essentially model--independent; it is both the
Goldstone boson and a bound state of strongly dressed
quarks~\cite{Maris:2003vk,Maris:1998hd}.
This reduction of QCD to an understandable case benefits us as the
circumstances of the lattice (that is equal mass quarks as discussed
later) may be investigated.  

The DSE approach of Tandy~\cite{Tandy:2003hn}
to the pseudoscalar decay
constant is presented in \Fig{fig:DSE} by the solid curve.
For the dashed curve, the same calculated decay constant is plotted
against the pseudoscalar mass obtained from $m_q$ via the
Gell-Mann--Oakes--Renner relation, as would
be used by \chpt{}.  This alternative approach
differs from \chpt{} and the lattice calculations as quark loops are
explicitly excluded from the calculation.  It would thus be expected
that the exclusion will, in the light quark mass region, result in a
behaviour similar to that expected by the JLQCD collaboration in
\Eq{eq:JLQCD} --- no logarithms, and yet be suggestive of the lattice
calculations in the region where quark loops become suppressed.  Tandy
parameterised the DSE results with the following phenomenologically
motivated form~\cite{Maris:1999nt}:

\begin{figure}[htb]%
  \includegraphics[angle=-90,width=78mm]{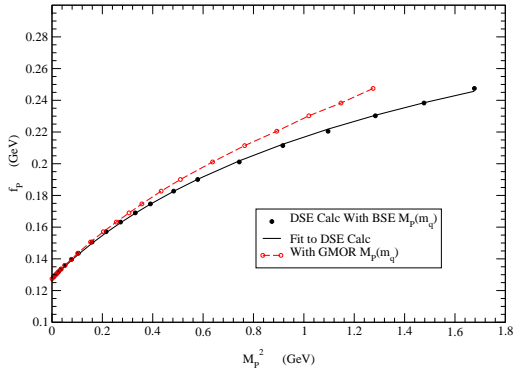}%
  \caption{The pseudoscalar decay constant versus the square of the
  pseudoscalar mass both calculated within the Dyson--Schwinger
  framework.  The solid line is the exact pseudoscalar mass, whilst
  the dashed curve is the mass obtained by using the
  Gell-Mann--Oakes--Renner relationship.~\cite{Tandy:2004privcomm}}
  \label{fig:DSE}
\end{figure}%

\be
\fPS = \sqrt{ \frac{f_{\pi}^{2}(0) + 
    A_{0}\mPS{}}{1.0 + A_{1}\mPS{} + 
    A_{2} \mPS{2}}}
\ee
Whilst reproducing the correct QCD behaviour at large $\mPS{}$ it is
incorrect at small quark masses.  As was discussed above this should
not be considered a failure, as the model that motivated this form
lacks the chiral behaviour that induces the logarithm in
\Eq{eq:ChPT}.

\subsection{Modified \chpt}
The final approach we will mention is a minimalistic modification of
\chpt{} as motivated by our previous work
\cite{Leinweber:1999ig,Leinweber:2001ac,Young:2002cj,Dunne:2001ip}.
In these investigations we found that a change in the form of the
regularisation of \chpt{} admits an extended radius of convergence
whilst maintaining an exact agreement with the results of the
dimensional regularised solution.  Recent work of the
Adelaide group~\cite{Young:2002ib} has show that this exact agreement
is indeed obtainable between different formulations of the
regulariser, and it is indeed a requirement of a regularisation
scheme.

The changes we make to the dimensional regularised result of \chpt{}
to form our Modified \chpt{} are two:
\begin{itemize}
  \item Terms of $\Order(\mPS{4})$ and higher in the full expression
  of \Eq{eq:ChPT} are neglected,
  \item An additional parameter $\X\mPS{2}$ is introduced to absorb
    the physics ignored by the truncation.
\end{itemize}
The form used for extrapolating thus becomes:
\bea
\frac{\fPS}{f_{\pi}(0)} & = &1 - 2 \left( 
      \frac{\mPS{}}{4\pi f_{\pi}(0)} \right)^{2}\non\\
      && \hspace{5mm}\times
      \log\left(\frac{\mPS{2}}{\X\mPS{2}+\Lambda^{2}}\right)
\label{eq:MChPT}
\eea
where the new term $\X\mPS{2}$ suppresses the chiral logarithms above a
certain energy scale and provides the missing analytic behaviour near
the chiral limit.  A similar approach has been attempted by
JLQCD~\cite{Aoki:2003xb}.

The advantages of this form for extrapolating results from lattice QCD
calculations are that it reproduces \chpt{} near the chiral limit, but
it also has the intermediate mass behaviour of QCD that the lattice
has shown.  This simple form allows an extension of the radius of
convergence of \chpt{}.  It must be noted that this form is indeed
na\"{\i}ve in the way the two mass limits are enforced but it does 
suggests hope for a more rigorous FRR solution that has been achieved
elsewhere~\cite{Young:2002ib}.  Such an extension is part of an
ongoing investigation.

\begin{figure*}[htb]%
\begin{center}
  \includegraphics[angle=90,width=101mm]{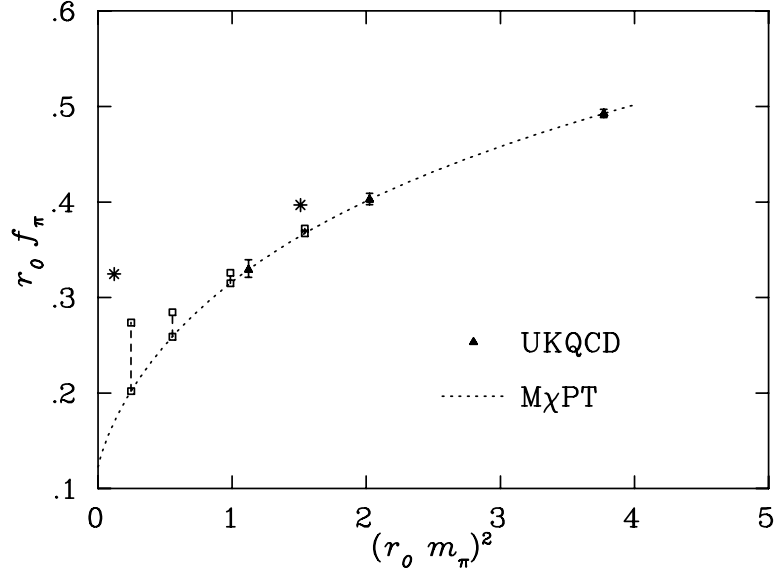}%
  \caption{The pseudoscalar decay constant in units of $r_0$ from
  UKQCD versus the squared pseudoscalar meson mass.  The experimental
  values of $f_{\pi}$ and $f_{K}$ are shown (*).  The dotted curve is
  the fit of the Modified \chpt{} result of \Eq{eq:MChPT}.  The
  vertical lines indicate expected finite size effects at various
  pseudoscalar meson masses, as in \Fig{fig:chlog}.}
  \label{fig:MChPT}
\end{center}
\end{figure*}%

This modified \chpt{} approach is more suited to extrapolating lattice
QCD calculations, at the current juncture, than the dimensionally
regularised approach of \Eq{eq:ChPT} for another reason too: it has
fewer fit parameters. As shown in \Fig{fig:ukvsjl} near the chiral
limit we currently have, at best, 5 data points, and fitting purely to
our UKQCD data we would have only three data points.  The
$\Order(p^{6})$ low--energy constants expansion of \Eq{eq:ChPT} (see
Ref.~\cite{Colangelo:2003hf} for the full expression) has seven
parameters: $f_{\pi}(0)$, $\Lambda_{1}$,~\ldots,~$\Lambda_{4}$,
$\tilde{r}^{F}(\mu)$ and $\mu$; each of which must be fit to some
data.  This has been undertaken in~\cite{Colangelo:2001df,Gasser:1984yg},
but in the comparison of lattice QCD to experimental results, these
parameters must independently be determined by the theory.

Conversely the simplistic approach of \Eq{eq:MChPT} has only three
parameters $f_{\pi}(0)$, $\Lambda$ and $\X$.  In comparing these two
approaches the expectation that the non-analytic behaviour should be
the same is an important success.  The values of $\Lambda$
(\Eq{eq:MChPT}) and $\Lambda_{4}$ (\Eq{eq:ChPT})
 are consistent within the large uncertainties as
previously discussed in~\cite{Leinweber:2001ac}.  The value of
$\Lambda_{4}$ from~\cite{Colangelo:2003hf} is $1.25
\stackrel{+0.15}{_{-0.13}}$ GeV whilst we find $\Lambda = 0.93 \pm
0.42$ GeV.  In a similar vein to Ref.~\cite{Dunne:2001ip} some of the
physics we have neglected, i.e. higher powers of $\mPS{}$, has
been absorbed into the parameter $\X$.  We present our fit of
\Eq{eq:MChPT} to our UKQCD data in \Fig{fig:MChPT}.  The vertical
dashed lines are the same finite size error estimation calculations as
discussed in Sec.~\ref{sec:comparison}.

It is clear that even with a finite size correction to the UKQCD
calculations there is a discrepancy between the data obtained from
lattice simulations and the experimentally measured decay constant of
the $K$ meson.  This discrepancy highlights some of the limitations of
the current results.  The finite size corrections that have been
applied to this data are indicative only and do not necessarily
account for all the finite size effects in the simulation.  The
benefit is that they reinforce the intuition that finite size
effects become less important as the quark masses get larger.

The discussion herein of an extrapolation of the quark mass ignores
the other important extrapolation that has not been undertaken in
these simulations; that is the continuum extrapolation.  In this work
we have chosen a value of $r_{0}$ as suggested by
\cite{Allton:2001sk,Dougall:2003hv}, however the uncertainty of this
value is 5\%.  Additionally, the value of $Z_{A}$ we use to
renormalise our result is determined perturbatively.  At first order
the correction is 25\% and the systematic error at second order is
expected to be up to 5\%. Whilst the value of $Z_{A}$ remains
determined perturbatively we would expect that part of the
discrepancy in our results to remain.

Finally there is a fundamental difference between the object that is
calculated at non-zero quark masses and the pseudoscalar meson that is
experimentally detected.  The $K$ meson conceptually is a
{\em strange} quark and a {\em light} quark in a sea of {\em light}
quarks.  The pseudoscalar meson calculated on the lattice with the
same mass consists of two quarks with masses half of that
of the strange quark in a sea of equivalently heavy quarks.  The DSE
calculations shown in \Fig{fig:DSE} have this behaviour, but the
experimental points of Figs.~\ref{fig:chlog} and \ref{fig:MChPT} are
different objects.  There is an expectation that the difference
between a $\langle qQ \rangle$ object and
$\langle \frac{Q}{2}\frac{Q}{2} \rangle$
meson is not large, but this is yet to be quantified.

\section{Conclusion}
We have presented a review of our recent publication
\cite{Allton:2004qq} and in particular the discussion of the
extraction of the pseudoscalar decay constant on the lattice.  We have
shown that this data set contains an indication of the onset of chiral
logarithms for the first time in this quantity.  Additionally we have
discussed how the need to extrapolate to the chiral limit still
exists and
furthermore that the suggestion of curvature necessitates the use of
chiral perturbation theory.
Finally we discuss some approaches and present a first attempt
at a form that respects chiral perturbation theory whilst being
applicable in the region where the lattice calculations are still
being performed.

\section*{Acknowledgements}
The author would like to acknowledge Peter Tandy not only for
providing \Fig{fig:DSE} but for the helpful discussions and Craig
Roberts for the same.
SVW would also like to thank Chris Michael and Alan Irving for
assistance during this research and helpful comments in the
preparation of this proceedings. The support of the U.K.~Particle
Physics and Astronomy Research Council is gratefully acknowledged.

\begingroup\raggedright\endgroup

\end{document}